\documentclass[12pt]{article}
\usepackage{scicite}
\usepackage{times}
\usepackage{graphicx}

\newenvironment{sciabstract}{%
\begin{quote} \bf}
{\end{quote}}

\newcounter{lastnote}

\title{Divestment may burst the carbon bubble if investors' beliefs tip to anticipating strong future climate policy}
\author{%
Birte Ewers,$^{1,2,\ast}$ Jonathan F. Donges,$^{1,3,\ast,\#}$ Jobst Heitzig$^{1}$, Sonja Peterson$^{4}$
\\
\normalsize{$^{1}$Potsdam Institute for Climate Impact Research, Member of the Leibniz Association},\\ 
\normalsize{P.O. Box 60 12 03, 14412 Potsdam, Germany}\\
\normalsize{$^{2}$Department of Economics, University of Kiel, Olshausenstra{\ss}e 40, 24098 Kiel, Germany}\\
\normalsize{$^{3}$Stockholm Resilience Centre, Stockholm University, Kr\"aftriket 2B, 114 19 Stockholm, Sweden}\\
\normalsize{$^{4}$Kiel Institute for the World Economy, Kiellinie 66, 24105 Kiel, Germany}
\\
\normalsize{$^\ast$The first two authors share the lead authorship.}\\
\normalsize{$^\#$To whom correspondence should be addressed; E-mail: donges@pik-potsdam.de}
}
\date{\today}

\begin{document} 
\baselineskip24pt
\maketitle 
\begin{sciabstract}
To achieve the ambitious aims of the Paris climate agreement, the majority of fossil-fuel reserves needs to remain underground. As current national government commitments to mitigate greenhouse gas emissions are insufficient by far, actors such as institutional and private investors and the social movement on divestment from fossil fuels could play an important role in putting pressure on national governments on the road to decarbonization. 
Using a stochastic agent-based model of co-evolving financial market and investors' beliefs about future climate policy on an adaptive social network, here we find that the dynamics of divestment from fossil fuels shows potential for social tipping away from a fossil-fuel based economy. Our results further suggest that socially responsible investors have leverage: a small share of 10--20\,\% of such moral investors is sufficient to initiate the burst of the carbon bubble, consistent with the Pareto Principle.
These findings demonstrate that divestment has potential for contributing to decarbonization alongside other social movements and policy instruments, particularly given the credible imminence of strong international climate policy. Our analysis also indicates the possible existence of a carbon bubble with potentially destabilizing effects to the economy.
\end{sciabstract}

\begin{figure}
\includegraphics[scale=1]{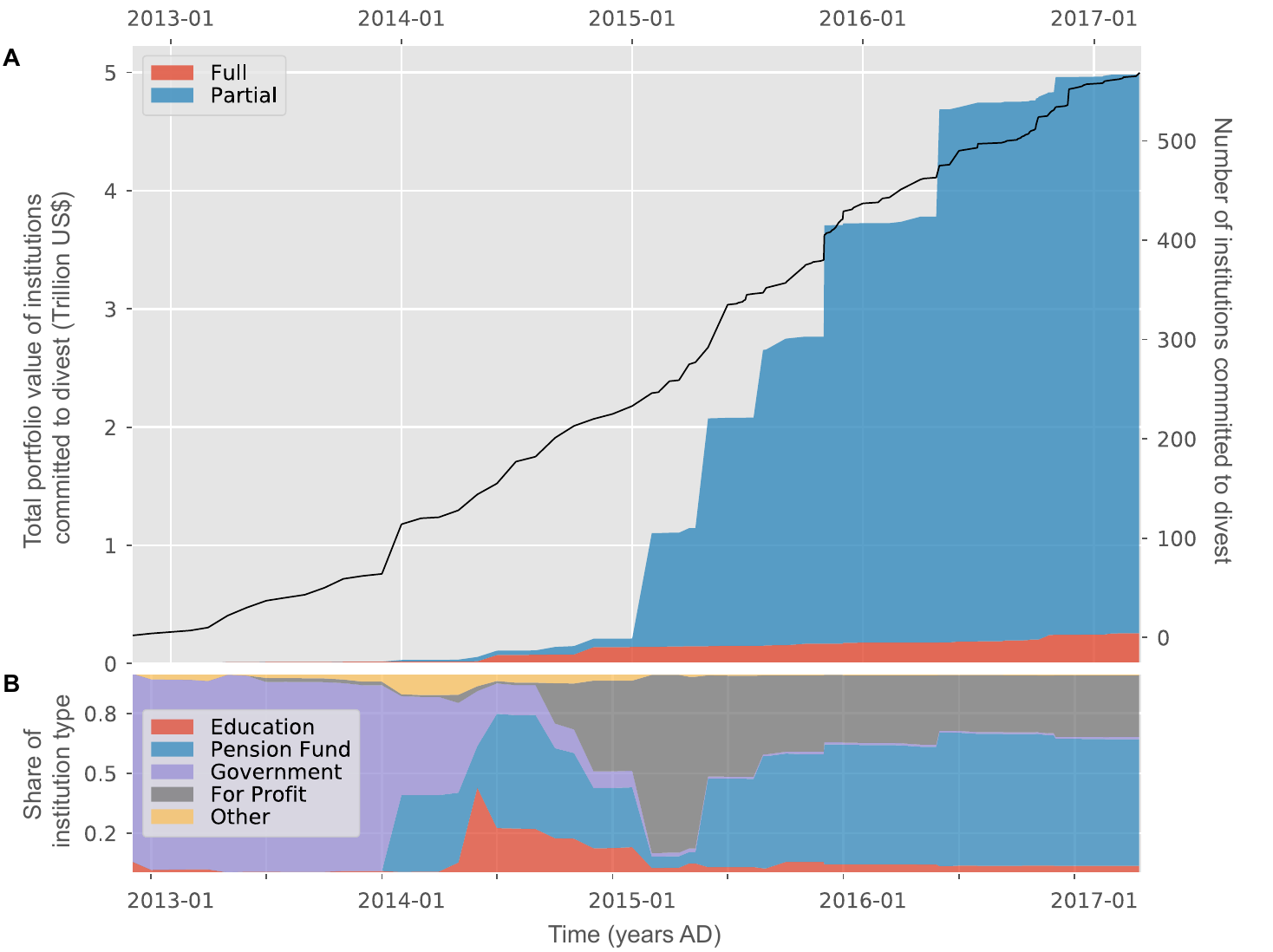}
\caption[Growth of divestment commitments]{
\textbf{Growth of institutional commitments to divest from fossil fuels.}
(A) Total portfolio value of institutions committed to divest (filled area, right-hand scale) and the number of institutions divesting (black line, left-hand scale) until March 2017. 
The value refers to the total portfolio value and not to the amount that is actually divested from the fossil fuel sector. 
The share of fossil fuel stocks in a portfolio, and thus the share to be divested, is typically 5--10\,\%. 
(B) Composition of the portfolio in share of total value by institution type over time.
Data provided by Fossil Free, a project of 350.org.
}\label{fig:timeline}
\end{figure}

With the Paris climate agreement signed~\cite{UNFCCC2015}, the world has made an important step towards embarking on rapid decarbonization of global socio-economic systems to mitigate anthropogenic climate change~\cite{rockstrom2017roadmap} with dangerous impacts on human societies and the biosphere~\cite{rockstrom2009safe,steffen2015planetary}. 
However, if all currently proven fossil fuel reserves were to be extracted, the resulting emissions alone would suffice to surmount the admissible carbon budget threefold \cite{McGlade2015}, amplifying the risk of runaway global warming~\cite{winkelmann2015combustion,steffen2018trajectories}. 
This apparent contradiction to the aim of climate change mitigation increasingly leads investors to see investments in fossil fuel assets as a moral issue, similarly to the divestment movement on investments in South Africa under the Apartheid regime starting in the 1970s~\cite{millar2018principles} or the English divestment movement pushing for the abolition of plantation slavery since the 17th century  \cite{Seidman2015}.
The same realization led to the creation of the Fossil Free movement in 2012,
urging investors on moral grounds to divest from fossil fuel companies \cite{McKibben2012}. 
To date, more than 1,000 institutions holding approximately \$\,7.93\,trn of assets have pledged to completely or partly divest from fossil fuels (Fig.\,\ref{fig:timeline}, data provided by Fossil Free, a project of 350.org \cite{FossilFreeData}), among them notable institutions such as the Rockefeller Brothers Fund and the Norwegian Sovereign Wealth Fund \cite{Mattauch2015}. 

At the same time, awareness of carbon risks, i.e., the risk of shares being devaluated and becoming stranded assets as fossil resources remain unburned, has been spreading \cite{ansar2013stranded,covington2016global,mercure2018macroeconomic}. 
Some believe that a carbon bubble is currently emerging that could burst when investors' expectations of carbon risk reach a critical threshold or when strict climate regulation such as carbon taxes and stringent emissions trading systems are enforced \cite{Gore2013,Leaton2014,Ritchie2015,bauer2018divestment}. 
It has been proposed that this process could contribute to rapid decarbonization of the World's socio-economic systems in accordance with the Paris agreement ~\cite{Schellnhuber2016,rockstrom2017roadmap}.
Recent analyses and empirical studies support this claim, showing that fossil fuel divestment can prevail over the green paradox~\cite{bauer2018divestment} and is able to mitigate financial risks posed by climate change and to reduce the carbon exposure of investment portfolios~\cite{battiston2017climate,hunt2018fossil}

Both processes, the moral and the risk-perception-driven divestment, could form a positive feedback leading to a self-organized social tipping dynamics \cite{Schellnhuber2009,milkoreit2018defining,tabara2018positive} of general investment behavior. 
And such tipping of social norms and conventions relevant for achieving sustainable development may be supported by parameters that can be influenced by policy makers \cite{Nyborg2016} as has been recently demonstrated by behavioral experiments~\cite{centola2018experimental} and in a model of social norms and social tipping points related to water conservation~\cite{castilla2017social}.
%
\section*{Model-based analysis of divestment dynamics and social tipping points}
To highlight the driving factors for the divestment movement to trigger the burst of a carbon bubble, 
and to analyze the thresholds, likelihood and timing of such a scenario,
we develop a stochastic agent-based model combining elements from the literature on social dynamics of investors \cite{Castellano2009}, 
complex adaptive social networks \cite{Holme2006,Gross2008}, 
and agent-based financial markets \cite{Hommes2006}. 
Following \cite{Farmer2009,farmer2015third}, we use agent-based modeling as it allows for accounting for adaptation, bounded rationality and heterogeneity in investment behavior \cite{Samanidou2007}. 
The model includes (see Methods for details)
(i) a representative public firm extracting and exploring fossil fuel resources, 
(ii) a stock market where shares of the fossil fuel firm as well as those of a non-fossil fuel firm are traded according to simple heuristic rules, 
(iii) an adaptive social investor network on which investors' beliefs about the imminence of strong carbon policy spread, 
(iv) cumulative carbon emissions as the environmental variable that represents the level of climate change effected, and
(v) a fixed carbon budget (of 250 GtCO$_2$) in line with the 2-degree target that may or may not be enforced by carbon policy. 

We distinguish neutral investors (NIs) that base their investment decision exclusively on expected economic returns from socially responsible investors (SRIs)\cite{Hong2009}. 
NIs withdraw their investment for economic reasons
if they are convinced that strong carbon policy will inhibit the extraction of a substantial share of the company's reserve and thus their investment will become unprofitable~\cite{battiston2017climate,hunt2018fossil}. They reinvest if the share price drops to the value the firm would have if carbon policy was implemented. Also, they can cease to believe in the credible implementation of carbon policy if their unconvinced neighbors earn higher returns.
In contrast, when convinced about carbon policy, SRIs support the divestment campaign by withdrawing funds immediately since they take the latter as a signal that society deems these assets harmful and neither consider their profitability any longer nor switch their opinion in the future.
We model the spreading of beliefs about the likelihood of carbon policy as a social learning process in an adaptive social network of investors \cite{Traulsen2010a,Wiedermann2015}. The central parameter describing this process is the typical time scale of social interactions, the social interaction frequency (SIF), described as the probability that an investor interacts with another investor randomly drawn from his/her social network neighborhood within a certain time to potentially spread his/her belief via a contagious-like process \cite{brockmann2013hidden,lehmann2018spreading}.

Our model is conservative in three respects:
First, we model only the spreading of beliefs about the likelihood of credible carbon policy and neglect the potential spreading of social norms that would turn NIs into additional SRIs, by assuming a fixed share (of 15\,\% default) which is in accordance with recent empirical findings for the US \cite{Foundation2014}.
Second, we highlight the importance of carbon policy
and its interaction with investors' decision-making 
by assuming that as long as SRI's do not believe in carbon policy,
they decide on purely economic grounds and may hold carbon assets if they appear profitable.
Third, we assume the implementation probability of the policy to grow slowly over time as more investors expect it to come, representing a feedback from public investors' opinion on carbon policy to actual policy processes. The probability only becomes significant when a vast majority is convinced that carbon policy will be implemented. 
\section*{Results}
\begin{figure}[htbp]
\includegraphics[scale=.95]{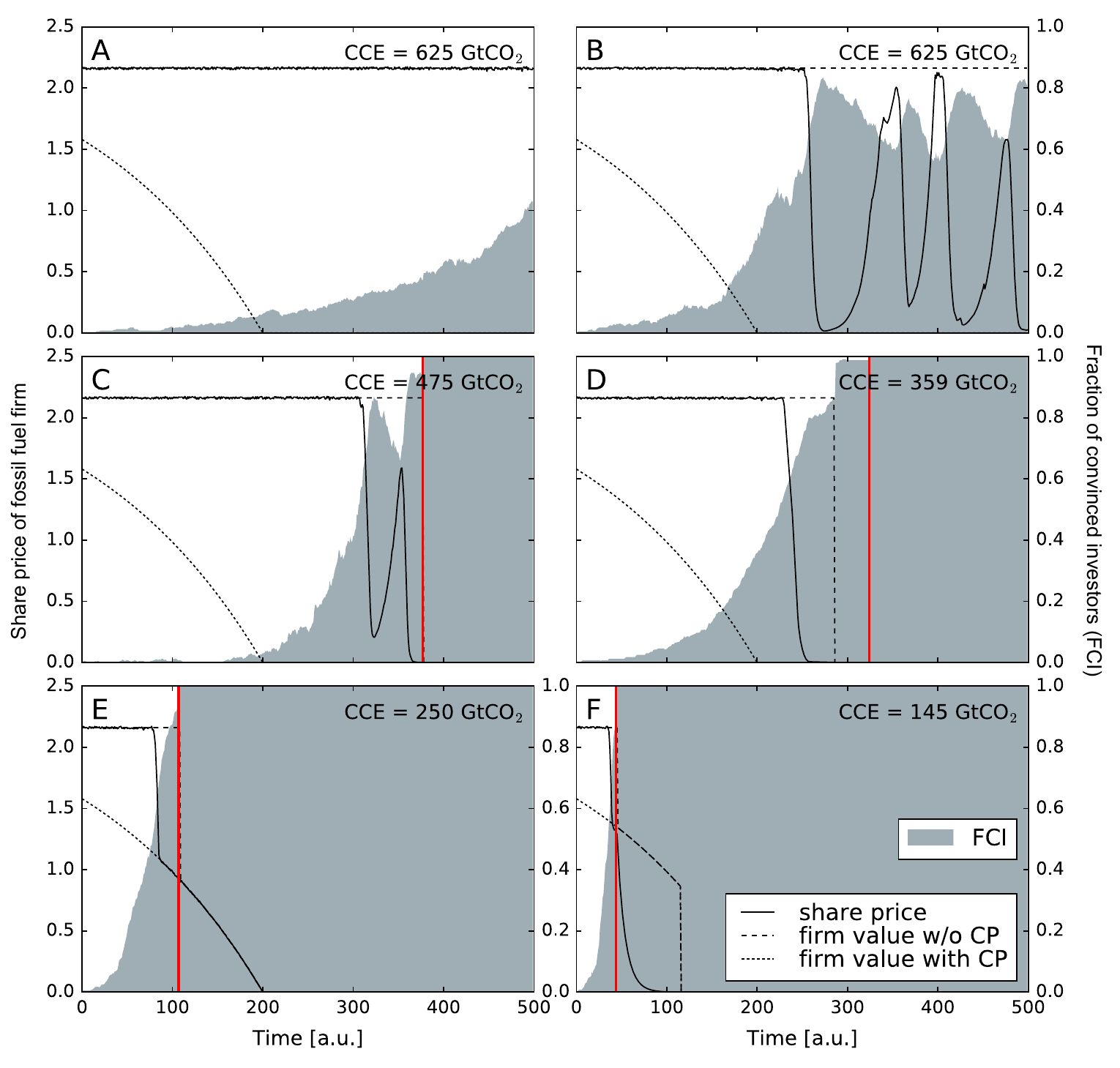}
\caption[Types of emergent behavior]{\textbf{Types of emergent behavior in modeled divestment from fossil fuels dynamics.} 
Representative evolution of the share price of a representative fossil fuel firm (solid line), the firm value without (w/o, dashed) and with (dotted) carbon policy (left scale),
and the fraction of investors convinced of the imminence of strong international climate policy (FCI, filled area, right scale) for different parameter values. 
The vertical red line indicates the time of implementation of carbon policy.
For each behavior type, the final level of cumulative carbon emissions (CCEs) is indicated.}
\label{fig:single-runs}
\end{figure}

%
We performed Monte Carlo simulations to generate an ensemble of possible time evolutions for three performance indicators, 
namely the share price of the representative fossil fuel firm, 
the fraction of ``convinced'' investors (FCI) who believe in carbon policy, 
and finally our central indicator of interest, the cumulative carbon emissions (CCE),
for a variety of parameter choices. 
Six dominant types of behavior emerge (Fig.\,\ref{fig:single-runs} A--F), 
corresponding to several metastable dynamical regimes. 
A first notable result is that initially the share price is virtually unaffected by a growing FCI in all types 
since withdrawn investments are replaced by NIs' funds as long as the market is sufficiently liquid.

In type (A), divestment is not successful since the belief in carbon policy spreads so slowly that 
(i) the share price is unaffected for much longer than it takes the firm to exhaust the carbon budget,
and (ii) the firm can do so since carbon policy is not implemented.

In type (B), divestment is partially successful. 
The belief in carbon policy spreads faster, so that at some point, 
too few unconvinced investors remain that would buy divested shares, 
and the share price drops without warning.
This can be interpreted as the bursting of a carbon bubble that grew as the belief in carbon policy spread. 
However, as the share price declines while the firm is still operating, the dividend per amount invested grows rapidly. 
As a consequence, convinced NIs switch their belief. 
Shares are bought and the share price rises
until the relative dividend has decreased again and the investment becomes less attractive
leading to any number of repetitions of this cycle.
Since the price never declines to zero and the policy is too unlikely to be implemented, 
the firm exceeds its carbon budget and CCE continue to rise. 

In type (C), divestment is also partially successful.
The first price drop occurs after the budget is exceeded but it is so severe that the firm is delisted and stops operating before the price can rise again, even without carbon policy. 
The latter is implemented only afterwards because the delisting quickly convinces the remaining investors and, hence, raises the probability for the policy.
But since CCE are already above the budget, we consider divestment only partially successful.

In type (D), the first price drop is still late, but carbon policy is implemented before the share price can rise again. 
At that point, all investors are convinced of carbon policy, 
evaluate the firm on the basis of its already negative remaining carbon budget, 
and, hence, divest immediately, so that the firm is delisted.
The firm stops extraction, freezing CCE at its current value.

In type (E), divestment is fully successful since the price drops before the carbon budget is exhausted.
The price stays at the value the firm would have 
if carbon policy was implemented and it could only extract its remaining carbon budget. 
This value is shown by the dotted black line (Fig.\,\ref{fig:single-runs}).
As the remaining budget approaches zero over time, the price follows, so that
the firm is delisted as soon as it has exhausted the budget.

In the final type (F), divestment leads to CCE below the carbon budget.
The share price drops below the firm value with carbon policy as the share of SRI's is very high.
Carbon policy is implemented before the budget is exhausted.

A common feature of types B--F and a central finding of this study is that 
the feedback between social dynamics of investors' beliefs on carbon policy and financial market dynamics 
can lead to the burst of a carbon bubble with share prices suddenly collapsing, 
when the belief in carbon constraints reaches a threshold value. 
 
\begin{figure}[htbp]
\begin{center}
\includegraphics[scale=1]{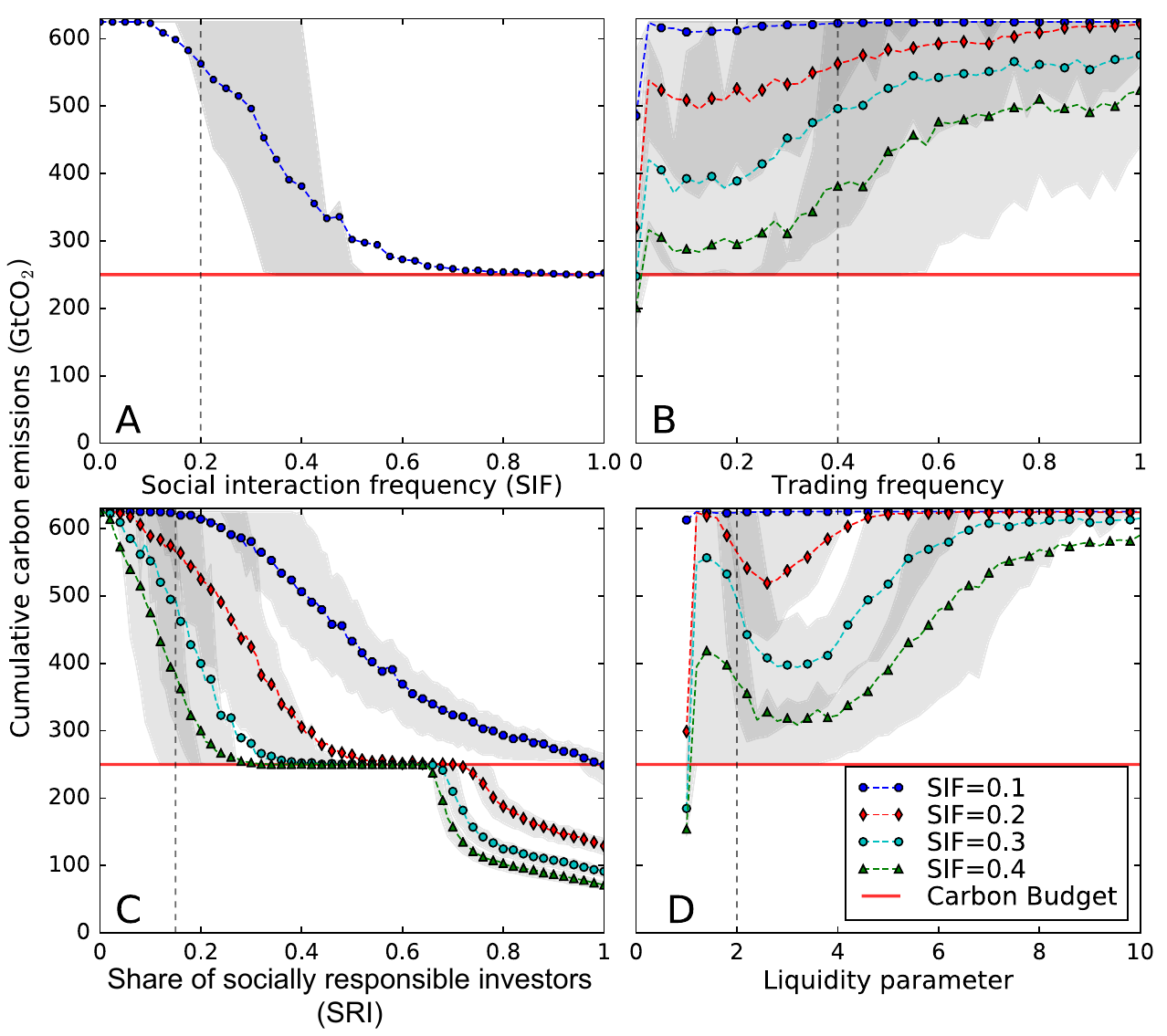}

\caption[Cumulative carbon emissions]{ 
\textbf{Resulting cumulative carbon emissions by social interaction and trading frequencies, share of social responsible investors (SRI), and initial wealth}
(mean value of 400 model runs each, see Methods for parameter values).
Overall, faster social interaction and a larger share of SRI-investors are beneficial for a successful divestment campaign while higher liquidity and faster trading tend to inhibit the burst of a carbon bubble.
Shaded region: interquartile range (first to third quartile). 
Dashed vertical lines: parameter's value used as baseline for other plots.
Horizontal line: carbon budget.
}
\label{fig:trading-parameters}
\end{center}
\end{figure}

We look at the influence of relevant key parameters on the relative frequency of these types of behavior and thus on the main performance indicator CCE (Fig.\,\ref{fig:trading-parameters}).
An important finding is that CCE is affected greatly 
not only by the share of SRIs and economic parameters such as the amount of wealth in the system and the trading frequency, 
but also by the parameters governing the social interaction.
In particular, the social interaction frequency of acquainted investors (SIF) has a considerable influence on CCE with a turning point and a greater variability existing within the interval $0.3$ and $0.4$ (Fig.\,\ref{fig:trading-parameters}A). 
For infrequent social interactions (small SIF), the vast majority of runs are of type A. 
For faster social interactions (larger SIF), the probability for type B--E to occur grows and the average CCE decline. 
Since the results are very sensitive to this parameter, 
the influence of other parameters is considered at different representative values of the SIF. 

The trading frequency, describing the rate at which an investor updates her investment decision, determines the typical time scale of the modeled economic dynamics. 
Faster trading has generally increases CCE 
with an exception for very small values at which investors react too slowly to changes in the FCI (Fig.\,\ref{fig:trading-parameters}B). 
For higher values of the trading frequency, a different mechanism dominates:
faster trading leads to prices reacting faster to disequilibria. 
For a low SIF of $0.1$ and a trading frequency larger than $0.4$, 
type A dominates and the average carbon emissions reach their maximum of 625\,GtCO$_2$.  

\begin{figure}[htbp]
\begin{center}
\includegraphics[scale=1]{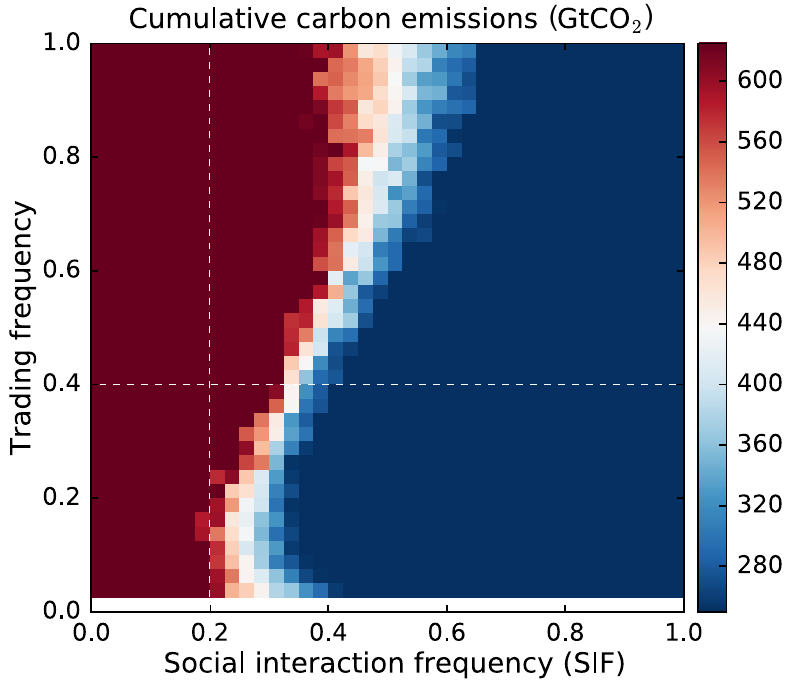}
\caption[Varying trading and social interaction frequency simultaneously]{%
\textbf{Complex tipping behavior and nonlinear interplay between effects of social interaction and trading frequencies on divestment dynamics.}
Cumulative carbon emissions resulting when varying social interaction and trading frequencies simultaneously (averages for 400 model runs each, baseline scenario, see Methods for parameter values). 
Horizontal and vertical lines: parameters' baseline values as used in Fig.~\ref{fig:trading-parameters}.
The white region indicates the location of the tipping point. 
\label{fig:cm-trading}
}
\end{center}
\end{figure}

Furthermore, we observe nonlinear feedbacks effect between SIF and trading frequency illustrating the importance of the relative size of the process time scales that dominate the modeled dynamics (Fig.\,\ref{fig:cm-trading}). 
Social tipping behavior is clearly visible in the white region, 
indicating the transition between a regime with an overwhelming majority of runs with high CCE, and a regime with an overwhelming majority of runs with low CCE.
Generally, the divestment movement can develop successfully in our model only if social interactions between investors happen more frequently than trading transactions. This finding establishes a relevant link to the debate on the destabilizing effects of algorithmic high-frequency trading~\cite{Leal2016}, although the trading processes represented in our model should be understood as strategic portfolio restructurings and not as potentially very fast micro-trading decisions.

Highlighting the important role of social norms in divestment, we find that an increasing share of SRIs strongly reduces CCE (Fig.\,\ref{fig:trading-parameters}C). 
One notable result is that even for low values $<$ 15\,\% of SRIs, divestment can lead to the burst of a carbon bubble so that the carbon budget is met (type E). 
A second striking feature is that for values of SRI larger than 60\,\%, type F starts to occur resulting in average CCE lower than the carbon budget. 
If the majority of investors is socially responsible, 
the dynamics spread much faster since divested investors do not switch back to their initial opinion. 
If the demand of NI's is too low, since their share is small, 
the price can fall to zero even before the carbon budget is exhausted. 

The liquidity parameter determines initial investors' wealth 
relative to the initial value of the fossil fuel stock and has a nonlinear effect on CCE (Fig.\,\ref{fig:trading-parameters}D). 
A value smaller than 1 leads to an immediate sudden price drop 
since divested funds cannot be substituted. 
The higher the liquidity parameter, the larger average CCE become. 
This can be explained by the effect that the price drops 
if the wealth of neutral investors is insufficient to buy all shares at the unconstrained share price. 
The higher the wealth of investors, the longer NIs can replace divested funds, 
and thus dynamics of types A and B from Fig.\,\ref{fig:single-runs} become more frequent.

\section*{Discussion}
%
%
Overall, our model analysis highlights that the divestment dynamics shows potential for social tipping and bursting of the carbon bubble 
before a carbon budget as implied by the Paris climate agreement is exceeded. 
How likely this outcome is depends on the model's parameters. 
With respect to the economic dynamics, 
we have shown that the results depend critically on 
the amount of initial wealth in the system relative to the initial value of the fossil fuel stock. 
More importantly, a faster social interaction rate and a higher share of socially responsible investors (SRIs) 
tend to `destabilize' the system, causing the share price of the fossil fuel firm to collapse. 
Notably, if social interaction between investors is fast, as could be assumed to hold today due to online communication, international air travel etc., 
a share of SRIs of only about 10 to 20\,\% 
(which approximately equals their actual current share in the US \cite{Foundation2014})
appears to be sufficient to trigger a price collapse. This is consistent with a generalized Pareto Principle~\cite{Pareto1969} stating that ``the decarbonization of the world will be led by a critical minority of key agents that advance transformative action"~\cite{Schellnhuber2016}.
As it is empirically observed already in the changing shares of different institutional investor types (Fig.\,\ref{fig:timeline}) and is reproduced by our model, 
the SRIs (such as governments and universities in the data) initiate divestment, but profit-oriented investors follow suit (such as pension funds and for-profit investors in the data).
In the real world, such a change in the general investment norm may thus be supported
by public policies that influence the identified crucial parameters of how many institutional investors divest from carbon-based assets, how fast stock portfolios may be restructured, and how often investors interact socially and communicate about their expectations and strategies.
In this sense, our study may be seen as supporting the theory of policy-driven social norm change outlined in \cite{Nyborg2016}.
These findings demonstrate that the divestment movement has potential for contributing to decarbonization alongside other instruments and social movements, particularly given the credible imminence of strong international climate policy. Our analysis also indicates the possible existence of a carbon bubble with potentially destabilizing effects to the economy.
\section*{Conclusions}
%
%
Summing up, we present an approach to modeling the divestment movement including economic and social interactions as well as a feedback on cumulative carbon emissions (CCEs). The model serves as a narrative for possible emerging qualitative dynamics and relies on simplifying assumptions such as a constant price for fossil fuels, identical wealth levels of investors and an uncorrelated price for fossil fuel stocks. Also, we assume a direct (supply-based) effect on CCE in response to a collapse in share prices omitting demand effects. In reality, divestment might affect CCE indirectly by triggering the implementation of carbon policy or by changing consumer behavior and thus lowering demand for fossil fuels. In addition, a large share of fossil fuel reserves is government/privately owned, and thus not traded on stock markets, which we do not consider in our model.

For the divestment movement, exciting times lie ahead. 
It has proven to have a reach beyond socially-responsible investors, 
and with the Paris climate agreement signed and ratified, it is likely that an increasing number of investors will start to incorporate carbon risks into their valuations.
The trend of declining oil prices over the past decade might be the first indication for a fundamental change in the fossil fuel industry, as carbon constraints become an undeniable fact on the path towards a transition to a carbon-neutral future.

\section*{Methods}

\subsection*{Data}

Data on the time evolution of institutional commitments to divest from fossil fuels from 2013--2017 has been provided by Fossil Free, a project of 350.org \cite{FossilFree2016} and is available from this organization by request.

\subsection*{Model description: the DIVEST model}

\subsubsection*{Representative fossil fuel firm}

In our model, a single representative, publicly owned, firm $F$ 
extracts fossil fuels at rate $q$ from a reserve $R_t$ that grows at an exploration rate $x$
and has initial value $R_0$.
When $R_t = 0$, the firm stops extracting and exploring and goes bankrupt.
Extracted fuels also reduce a theoretical remaining carbon budget $B_t$,
which is initially $B_0$.
At each discrete time point $t$ [a.u.], a carbon policy may come into force which makes the carbon budget binding.
If it is in force and $B_t \le 0$, the firm also goes bankrupt.

\subsubsection*{Investor types and evaluation of shares}

There are $I$ investors with monthly discount rate $r$ (months are used to denote model time steps below),
some share of which is ``socially responsible'' (coded as $\gamma_i=1$, the others as $\gamma_i=0$).
At each $t$, each investor $i$ either does ($\beta_i(t)=1$) or does not ($\beta_i(t)=0$) believe that the carbon policy will come into force.
If they don't and $F$ is not bankrupt, 
they evaluate a share of $F$ at its unconstrained net present value
$NPV^u_t = DPS_f\times(1-e^{-r R_t/q})/r$,
where $DPS_f = q p / N$ is the expected dividend per share, 
$p$ the net price, i.e. the market price for fossil fuels minus extraction and exploration costs
and $N$ the number of shares issued.
If they do believe in carbon policy coming into force, they instead use the smaller net present value constrained by the carbon budget
$NPV^c_t = DPS\times(1-e^{-r B_t/q})/r$.

\subsubsection*{Economic dynamics on stock market}

At each time point, 
each investor $i$ has their wealth invested into either the fossil firm or an alternative asset that provides dividends fluctuating slightly around a fixed mean value at random.
Each month, a share $\rho$ (called the ``trading frequency'') of investors update their investment decision, 
in random sequential order, 
based on the difference between their assessment of the fossil firm's value, $NPV_i$, 
and its current share price $s$ on a stock market.
If $NPV_i > s$ and not $\beta_i=\gamma_i=1$, they shift their complete investment to the fossil firm, 
otherwise to the alternative asset.
In particular, socially responsible investors who believe in climate policy divest no matter the price. 
A market maker updates the prices after each individual investor's decision in proportion to the change in the number of shares hold by investors \cite{LeBaron2001}.

\subsubsection*{Social dynamics of investors: social learning of beliefs about climate policy and homophilic social network adaptation}

Investors adjust their beliefs about climate policy via social learning through pairwise interaction on an adaptive, initially small world, network.
Each month, a share $SIF$ (called ``social interaction frequency'') of investors select a random network neighbor $j$ and interact with that person if their beliefs $\beta_i,\beta_j$ differ.
If so, $i$ either breaks the link to $j$ and connects instead with a randomly chosen investor $k$ with the same belief ($\beta_k=\beta_i$), which happens with probability $\phi=0.1$ (as in \cite{Holme2006}), 
or else considers adopting $j$'s belief with a probability depending on their respective successes (similar to \cite{Traulsen2010a,Wiedermann2015}).
They measure success as 
$\sigma_i = 100 ROI_i + WD_i / \overline{WD}$,
combining the current return on their total investment, $ROI_i$,
with the ratio of their current wealth and dividends, $WD_i$, to its current average $\overline{WD}$.
$i$ adopts the belief $\beta_j$ either with probability
$p_0 = (1+\tanh(\alpha(\sigma_j-\sigma_i)))/2$ (if $\gamma_j=0$ or $\beta_j=0$) 
or $p_1 = \min\{p_0 + \delta, 1\}$ 
(if $\gamma_j=\beta_j=1$).
In this,
$\delta > 0$ represents the fact that socially responsible investors who divested are assumed to be more convincing,
and $\alpha$ controls the influence of success on opinion dynamics.

\subsubsection*{Implementation of strong climate policy}

Finally, each month the anticipated climate policy may actually come into force, which happens with a probability of 
$p_c=\exp(-\lambda |\{i:\beta(i)=0\}| / N)$,
converging late but steeply to $p_c=0.1$ as the share of believers approaches 100\,\%,
where $\lambda=20$ controls this convergence.

\subsubsection*{Parameter baseline values}

Baseline values for the simulation are $R_0 = 500$, $B_0 = 250$, $r=0.005$, $q = x = 2.5$, $p=10$, $I=400$ and $N=1000$.

\subsection*{Code availability}

The Netlogo code describing the DIVEST model presented in this paper will be made available to editors and referees upon request. It will be released in an open repository such as github upon publication of the paper.

\section*{Author contributions}

B.E., J.F.D., J.H. and S.P. designed research; B.E. performed research and analyzed data; and B.E., J.F.D., J.H. and S.P. wrote the paper.

\section*{Competing interests}

The authors declare no competing interests.

\section*{Acknowledgements}

This work was developed in the context of the COPAN collaboration on complex human-environment systems in the Anthropocene at the Potsdam Institute for Climate Impact Research (www.pik-potsdam.de/copan). BE is grateful for financial support from the German National Academic Foundation (Studienstiftung des deutschen Volkes).
JFD thanks the Stordalen Foundation (via the Planetary Boundaries Research Network PB.net), the Earth League's EarthDoc program, the Leibniz Association (project DominoES) and the European Research Council advanced grant project ERA (Earth Resilience in the Anthropocene) for financial support. 
We thank HJ Schellnhuber, JD Farmer, C Hepburn, W Lucht, N Bauer and K Lessmann for helpful discussions.
We are grateful to Fossil Free, a project of 350.org, for providing the data that is the basis of Fig.~\ref{fig:timeline}.

\bibliography{scibib}

\bibliographystyle{naturemag}


\clearpage

\end{document}